\documentclass[12pt]{elsarticle}

\journal{}  
\makeatletter
\def\ps@pprintTitle{%
   \let\@oddhead\@empty
   \let\@evenhead\@empty
   \let\@oddfoot\@empty
   \let\@evenfoot\@empty
}
\makeatother

\usepackage{times}
\usepackage{amsfonts}
\usepackage{amsmath}
\usepackage{amssymb}
\usepackage{natbib}
\usepackage{graphicx}
\usepackage{enumitem}
\usepackage{xcolor}
\usepackage{subcaption}
\usepackage[normalem]{ulem}
\usepackage[colorlinks=true,linkcolor=red, citecolor=blue]{hyperref}
\begin{document}
\begin{frontmatter}
\date{}
\title{Does the Amati Correlation Exhibit Redshift-Driven Heterogeneity in Long GRBs?}
    \author[label1,label2]{Darshan Singh}
\affiliation[label1]{organization={Department of Basic and Applied Sciences},
            addressline={GD Goenka University},
             city={Gurugram},
             postcode={122103},
             state={Haryana},
             country={India}}
 \affiliation[label2]{organization={Department of Physics},
            addressline={Mata Raj Kaur Institute of Engineering and Technology},
             city={Rewari},
             postcode={123401},
             state={Haryana},
             country={India}}
 \author[label3]{Meghendra Singh} 
\affiliation[label3]{organization={Delhi Metro Rail Corporation Limited},
             city={New Delhi},
             postcode={110001},
             state={Delhi},
             country={India}}
\author[label1]{Dinkar Verma} 
\author[label4]{Kanhaiya Lal Pandey}
\affiliation[label4]{organization={Department of Physics, School of Advanced Sciences},
            addressline={Vellore Institute of Technology},
             city={Vellore},
             postcode={632014},
             state={Tamil Nadu},
             country={India}}
\author[label1]{Shashikant Gupta\corref{cor1}}
\cortext[cor1]{Corresponding author
}
\begin{abstract}
Long gamma-ray bursts (GRBs) offer significant insights into cosmology due to their high energy emissions and the potential to probe the early universe. The Amati relation, which links the intrinsic peak energy to the isotropic energy, is crucial for understanding their cosmological applications. This study investigates the redshift-driven heterogeneity of the Amati correlation in long GRBs. We analyzed 221 long GRBs with redshifts from 0.034 to 8.2, dividing them by redshift thresholds of 1.5 and 2.
Using Bayesian marginalization and Reichart's likelihood approach, we found significant differences in the Amati parameters between low and high redshift subgroups. These variations, differing by approximately $2\sigma$ at $z = 1.5$ and more than $1\sigma$ at $z = 2$, suggest an evolution in the GRB population with redshift, possibly reflecting changes in host galaxy properties. However, selection effects and instrumental biases may also contribute to it. Our results challenge the assumption of the Amati relation's universality and underscore the need for larger datasets and more precise measurements from upcoming missions like Transient High-Energy Sky and Early Universe Surveyor (THESEUS), and enhanced X-ray Timing and Polarimetry mission (eXTP) to refine our understanding of GRB physics.
\end{abstract}



\begin{keyword}
Cosmology, Gamma ray burst, Large-scale structure of the universe, Galaxies.

\end{keyword}

\end{frontmatter}



\section{Introduction}
\label{sec:intro}
Gamma-ray bursts (GRBs), observed as intense flashes of gamma rays in distant galaxies, represent some of the most energetic explosions in the universe \citep{Band1993,kumar2014}. A key characteristic of GRBs is their pulse duration, commonly quantified by the parameter $T_{90}$ \citep{kou1993}, defined as the time interval during which 90\% of the total background-subtracted gamma-ray photons are detected. Based on $T_{90}$, GRBs are classified into two main types: short GRBs, lasting less than 2 seconds and typically resulting from the merger of compact objects like neutron stars or black holes; and long GRBs, lasting more than 2 seconds and believed to be associated with the collapse of massive stars into black holes \citep{kumar2014, Amati2021}. Long GRBs are particularly valuable for cosmological studies due to their longer pulse duration, facilitating localization within their host galaxies and redshift measurement. They can help measure cosmological distances, probe the early universe, and provide constraints on cosmic reionization and star formation rates. 
Correlations among various GRB observables are important to calibrate the observational data for distance measurement. 
Among the various correlations observed, the Amati relation \citep{ama02,ama06,ama08,ama09}, which links the intrinsic peak energy ($E_{p,i}$) to its equivalent isotropic energy ($E_{iso}$), has been a subject of particular interest. $E_{p,i}$ is computed from the observed peak energy $E_{peak}$ in the $\nu F_{\nu}$ spectrum of a burst as $E_{p,i}=E_{peak}(1+z)$. 
Originally expressed in exponential form, the Amati relation is now commonly presented in linear form for its elegance:
\begin{equation}
 \log E_{iso} = a + b \log E_{p,i}  \, ,
\label{eq:amati} 
\end{equation}
where the intercept, $a$ and the slope, $b$ are known as Amati parameters. The parameter $b$ is usually positive indicating a positive correlation between $E_{iso}$ and $E_{p, i}$ while $a$ is negative. 
Alongside the Yonetoku relation between $E_{peak}$ and the isotropic peak luminosity, $L_{iso}$ \citep{Yonetoku2004}; Ghirlanda relation between $E_{peak}$ and collimation-corrected energy, $E_{\gamma}$  \citep{ghi04,ghirlanda2010}; and Dianotti relation \citep{Dainotti_2022}; Amati relation serves as a cornerstone for probing the physics of GRBs and their use as cosmological tools \citep{daigne2006,basak2013,dai2021,huang2021}. 
Several studies have attempted to constrain cosmological parameters using GRB data \citep{lin2016model, Demianski17, Cao_2021, Khadka2021,  jia2022, 2024arXiv240519953C}. However, a critical question arises regarding whether the relations used for cosmological applications remain constant throughout the history of the universe or evolve with redshift, a measure of the cosmic scale and the age of the universe at the time of the burst \citep{Dainotti_2013}. GRBs are calibrated using type Ia supernovae for cosmological applications, which can be observed only up to $z=1.75$. If high-redshift GRBs differ from their low-redshift counterparts, their calibration may be questionable.
Recent studies have presented conflicting views on the evolution of the Amati parameters with redshift. Some research suggests that the Amati parameters, $a$ and $b$, may systematically vary with the mean redshift of GRBs. By dividing a sample of long-duration GRBs into redshift-based groups and fitting the Amati relation separately to each group, significant and systematic changes in the parameters with redshift have been observed \citep{dai2021, huang2021, jia2022, Li2007, Geng_2013, Tsutsui_2013, lin2015long, Singh2024}. Monte Carlo simulations further support that this variation is unlikely due to selection effects from the fluence limit, indicating a strong evolution of GRBs with cosmological redshift \citep{Li2007}. In a recent study \cite{kumar2023gamma}, it is concluded that the Amati relation of GRBs evolves with $z$. 
Conversely, other studies \citep{Butler_2009, zitouni14} challenge this view by using different samples of GRBs and investigating the redshift independence of the Amati and Yonetoku relations. By binning the data by redshift and fitting both relations, these studies found that the normalization and slope do not exhibit systematic evolution with redshift, implying that the Amati and Yonetoku relations are redshift-independent.
The discrepancy between these findings raises important questions about the intrinsic properties of GRBs and their use as standard candles in cosmology.

Differences in the properties of GRBs at different redshifts, if exist, may reflect signatures of cosmic evolution. For instance, the peak of star formation in cosmic history at $z= 1.5-2$ could be detected in GRB properties. Substantial evidence indicates evolutionary shifts in the demographics of various galaxy morphologies with increasing redshift. Notably, there is a significant decline in the proportion of disk galaxies around $z=2$ \cite{mortlock2013redshift} accompanied by a corresponding increase in peculiar types. However, for massive galaxies, the fractions of disks, spheroids, and peculiar types appear to remain relatively constant within the redshift range of $1.5 < z < 6.5$. Furthermore, galaxy morphologies exhibit limited correlation with other key physical properties such as star formation rate, color, mass, or size \citep{mortlock13,conselice14,ferreira2022}. Mass distribution of host galaxies and the redshift distribution of long GRBs, as investigated by \citep{Wang2014} indicates that GRB host galaxies are metallically biased tracers of star formation. At high redshift, the early-type galaxies can be observed in the active star formation period, during which massive star explosions and galactic winds occur. This situation is not found at lower redshift, where star formation is already quenched for this morphological type \citep{Palla2020} which might be linked to a low GRB rate.
Contrary to this, recent observations by the James Webb Space Telescope (JWST) have identified massive galaxies at high redshifts, posing a challenge to the standard $\Lambda$CDM cosmological model \citep{2023Natur.616..266L}.

To address this knowledge gap, i.e., whether there are two different populations of long GRBs at low and high redshifts, this study investigates the differences in Amati parameters in the low and high-$z$ regimes. In Section~\ref{sec:data-method}, we will describe our methodology, including the data and the analysis techniques employed. Section~\ref{sec:resndisc} presents the key findings of our research and discusses the implications of these findings.  Finally, Section~\ref{sec:conclusion} will conclude the paper by summarizing the main points and suggesting avenues for future research.
\section{Data and Methodology}
\label{sec:data-method}
\subsection{GRB Data Sample}
\label{sec:data}
Investigating the evolution of GRBs with redshift requires large datasets of long GRBs covering a wide range of redshifts. In a previous study, \citep{Singh2024} used 162 long GRBs compiled by \citep{Demianski17} to explore differences between low and high-redshift GRB populations. However, \citep{Dirirsa2019} and \citep{Khadka2021} have raised concerns about the suitability of this dataset for cosmological applications, noting that the constraints derived from these GRBs show inconsistencies with those obtained from other cosmological probes such as Type Ia supernovae, BAO, H(z) measurements, and CMB anisotropy data. To address these concerns and validate the results of \citep{Singh2024}, a larger dataset with more accurate measurements of GRB observables is required. In the current study, we utilize a more comprehensive and recent dataset compiled by \citep{jia2022}, containing 221 long GRBs spanning a redshift range of $0.034 \leq z \leq 8.2$. While many GRBs in this new dataset overlap with the previous 162-GRB sample, they benefit from updated values of key parameters, including isotropic equivalent energy and peak energy. This dataset primarily derives from the joint observations conducted by Swift and Fermi satellites. 

In some cases, $E_{p,i}$ is provided directly by Swift which has significantly contributed to detecting several GRBs with their redshifts. Swift is equipped with three instruments: the Burst Alert Telescope (BAT) for GRB detection, the X-ray Telescope (XRT), and the Ultra-Violet and Optical Telescope (UVOT). BAT, with its large field of view, detects GRBs in the energy range of 15 keV to 150 keV. Upon detection, the satellite slews to the burst direction for XRT and UVOT observations. The energy range of XRT is 0.2-10 keV, while the UVOT is capable of observations in the $170-600$ nm wavelength range \citep{Gehrels2004}.
The Fermi Gamma-Ray Space Telescope (FGST), launched by NASA in 2008, has become a cornerstone in the study of high-energy gamma rays \citep{Atwood2009}. The primary instrument onboard FGST is the Large Area Telescope (LAT), a high-energy gamma-ray imaging telescope with a wide field of view, operating in the energy range from below 20 MeV to more than 300 GeV. Additionally, Fermi is equipped with the Gamma-ray Burst Monitor (GBM) and the Anticoincidence Detector (ACD), which complement the observations made by the LAT. The LAT provides unprecedented sensitivity across a broad energy range, spanning approximately 20 MeV to 300 GeV, making Fermi an invaluable tool in high-energy astrophysics.
\begin{table}
\begin{center}
\caption{Description of GRB data (G) and its subgroups. Groups $\rm{G_1}$ and $\rm{G_2}$ belong to the division at $z=1.5$, while $\rm{G'_1}$ and $\rm{G'_2}$ belong to the division at $z=2$. }
\label{tbl:data} \bigskip 
\begin{tabular}{ccccc}
\hline
 Group(s) & Number of GRBs & $z_{min}$ & $z_{max}$ & Median $z$\\
\hline
$\rm{G}$  & $221$ & $0.034$ & $8.20$ & $1.619$\\
\hline
$\rm{G_1}$ & $100$ & $0.034$ & $1.489$ & $0.893$\\
$\rm{G_2}$ & $121$ & $1.517$ & $8.20$ & $2.452$ \\
\hline
$\rm{G'_1}$ & $132$ & $0.034$ & $1.98$ & $1.10$ \\
$\rm{G'_2}$ & $89$ & $2.006$ & $8.20$ & $2.77$ \\
\hline\hline
\end{tabular}
\end{center}
\end{table}
\subsection{Methodology}
\label{sec:method}
Below we discuss the technique used for our analysis, A part of it has been presented in \cite{Singh2024}.

\subsubsection{Computing best-fit of Amati Parameters}
The observed values of the peak energy ($E_\mathrm{p,i}$) and isotropic-equivalent energy ($E_\mathrm{iso}$) along with their uncertainties, are available for all the GRBs in the data set. For the linear form of the Amati relation, i.e, Eq.~\ref{eq:amati} among the logarithm of the observational quantities, we define $\chi^2$ as:
\begin{equation}
\chi^2 = \sum_{i=1}^{N} \left( \frac{g^{i} -  f(E_{p,i}^i;a,b)}{\sigma^i} \right)^2 \, ,
\label{eq:chisq}
\end{equation}
where $g^i=\log E_\mathrm{iso}^i$, is obtained from data, and $f$ is the value of $\log E_\mathrm{iso}$ derived using Amati relation, i.e., corresponding to given $E_{p,i}^i$ (see Eq.~\ref{eq:amati}). 
The uncertainty $\sigma_i$ in $\log E_\mathrm{iso}^i$ can be calculated as described in \citep{Singh2024}:
\begin{equation}
    \sigma_i^2   = (dE_{\mathrm{iso}}/E_{\mathrm{iso}})^2 + b^2 (dE_{\mathrm{p,i}}/E_{\mathrm{p,i}})^2 \, . 
\end{equation}
The likelihood, $P(D|M)$, which represents the probability of obtaining the data for the given model $M$ can be expressed as:
\begin{equation}
P(D|M(a,b)) \propto e^{-\chi^2/2} \,  , 
\label{eq:likeli}
\end{equation}
where $\chi^2$ is defined by Eq.~\ref{eq:chisq}. The best-fit values of the Amati parameters ($a$ and $b$) can be estimated by minimizing $\chi^2$ in Eq.~\ref{eq:chisq} or by maximizing likelihood in Eq.~\ref{eq:likeli}. However, this method does not provide direct probabilities of these parameters, thus we employ the Bayesian approach. The direct probability of the model $M$, also known as the posterior probability can be easily calculated using Bayes' theorem :
\begin{equation}
     P(M(a,b)|D) \propto P(D|M(a,b)) P(M(a,b)) \, .
    \label{eq:bayes}
\end{equation}
Here, $P(M(a,b))$ represents the prior probability of the model. One should be cautious while selecting the prior probability as it can include personal biases and lead to an inappropriate posterior probability distribution. 
An advantage of the Bayesian approach is the ability to marginalize the undesired parameters. For example, marginalization on parameter $b$ can be performed by integrating over it:
\begin{equation}
    P({a}| I) = \int {P(I| a,b)P(a,b)db} \, .
    \label{eq:margin}
\end{equation}
To determine whether the Amati parameters systematically increase or decrease with redshift, we fit the Amati relation in Eq.~\ref{eq:amati} for each subgroup of data using the aforementioned methods. 
\subsubsection{Intrinsic Scatter In the Amati relation}
\label{sec:Intrinsic_Scatter}
The correlation between $E_p$ and $E_{iso}$ exhibits variability often attributed to inherent scatter \citep{Singh2024, Reichart_2001}. To mitigate this variability, we employ Reichart's likelihood function, which incorporates the intrinsic scatter ($\sigma_{int}$) along with the linear relation described by Eq. \ref{eq:amati} and is given as
\begin{eqnarray}
\label{eq:reichart}
L_{Reichart}(a, b, \sigma_{int}) & = & \frac{1}{2} \frac{\sum{\log{(\sigma_{int}^2 + \sigma_{y_i}^2 + a^2
\sigma_{x_i}^2)}}}{\log{(1+a^2)}} \\ \nonumber &&+ \frac{1}{2} \sum{\frac{(y_i - a x_i - b)^2}{\sigma_{int}^2 + \sigma_{y_i}^2 + a^2
\sigma_{x_i}^2}}\,.
\end{eqnarray}

The correlation function, $L_{Reichart}(a, b, \sigma_{int})$, now involves three parameters. These parameters can be simultaneously fitted, or the parameter $b$ can be analytically evaluated by setting $\displaystyle
{\frac{\partial }{\partial b}L(a, b, \sigma_{int})=0}$. This yields:
\begin{equation}
b = \left [ \sum{\frac{y_i - a x_i}{\sigma_{int}^2 + \sigma_{y_i}^2
+ a^2 \sigma_{x_i}^2}} \right ] \left [\sum{\frac{1}{\sigma_{int}^2 + \sigma_{{y_i}}^2 + a^2 \sigma_{x_i}^2}} \right ]^{-1}\,. 
\label{eq:calca}
\end{equation} 

To comprehend the impact of intrinsic scatter, we concurrently determine the optimal values for parameters $a$, $b$, and $\sigma_{int}$ using the aforementioned three-parameter likelihood function defined by Eq.~\ref{eq:reichart}. During the computation of the optimal value for one of the parameters, Bayesian marginalization is applied to the remaining two parameters. The same approach is applied to the different subgroups of the data as well.

\subsubsection{Separate Fits to Low and High-$z$ GRBs}
\label{sec:division}
As discussed in Section~\ref{sec:intro}, galaxy populations exhibit significant differences in metallicity below and above a critical redshift, ranging from approximately $1.5$ to $2$. The GRB production rate depends on the metallicity of the host galaxy \citep{Perley2016}. To explore the potential differences between GRBs below and above the critical redshift, we aim to examine the variations in the Amati parameters with redshift. For this purpose, we divide the data sample (hereafter $G$), containing 221 GRBs, into subgroups of low and high redshift.
When the critical redshift is set at $z=1.5$, the low-$z$ sample consists of 100 GRBs, with a median redshift of $0.893$ and a redshift range of $0.034 < z < 1.489$. The high-$z$ sample comprises 121 GRBs, within a redshift range of $1.517 < z < 8.2$ and a median redshift of $2.452$. These subgroups are referred to as $\rm{G_1}$ and $\rm{G_2}$, respectively.
When the critical redshift is set at $z=2$, the low-$z$ sample includes 132 GRBs, with a median redshift of $1.10$ and a redshift range of $0.034 < z < 1.98$. The high-$z$ sample consists of 89 GRBs, within a redshift range of $2.006 < z < 8.2$ and a median redshift of $2.77$. These subgroups are referred to as $\rm{G'_1}$ and $\rm{G'_2}$, respectively.

\begin{figure}
\centering
\includegraphics[width=.75\linewidth]{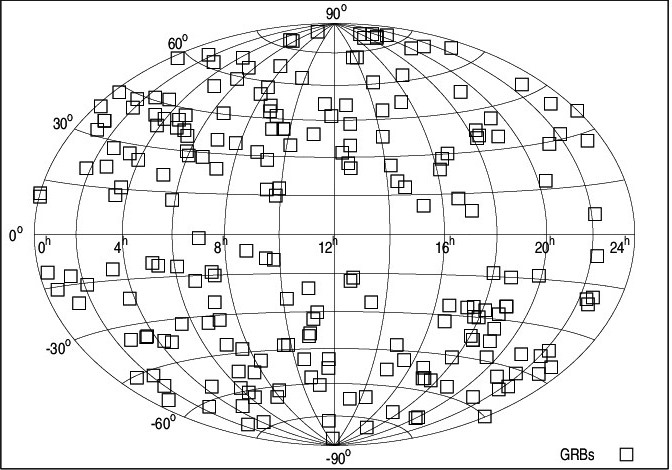}
\caption{This celestial map shows the isotropic distribution of 221 gamma-ray bursts (GRBs) across the sky.} 
\label{fig:skycoverage1}
\end{figure}

\begin{figure*}
    \begin{minipage}[t]{.50\linewidth}
    \includegraphics[width=\linewidth]{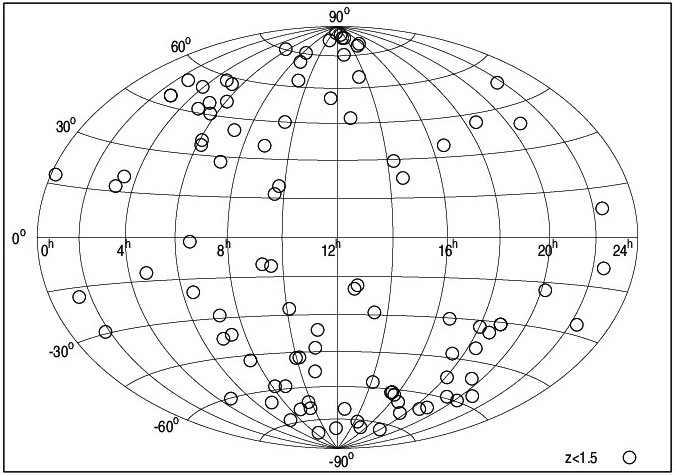}%
    \subcaption{$z<1.5$}
  \end{minipage}\hfil
  \begin{minipage}[t]{.50\linewidth}
 \includegraphics[width=\linewidth]{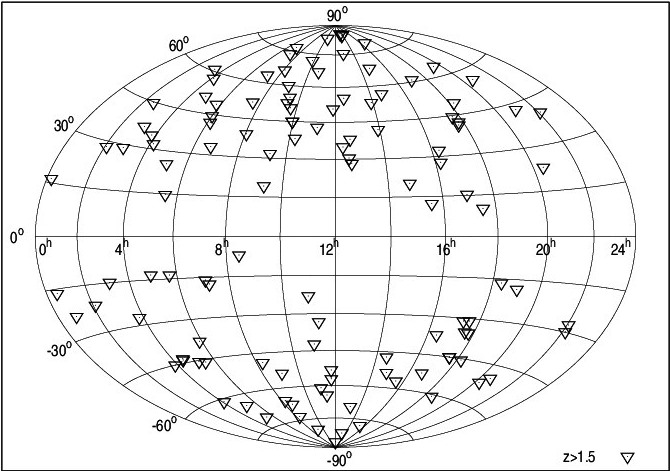}%
    \subcaption{$z>1.5$}
  \end{minipage}%
  \caption{This celestial map shows the isotropic distribution of low and high-$z$ gamma-ray bursts (GRBs) across the sky to ensure the fair sky coverage of the subgroups.}
  \label{fig:skycoverage2}
\end{figure*}

GRB observations indicate a statistically isotropic distribution of their positions \citep{balazs1998anisotropy,andrade2019}. A large number of GRBs is required to further investigate any potential 
deviations from isotropy \citep{Sethi2000}. To avoid any directional biases, we aim to ensure that our GRB sample and its subgroups adequately cover the sky. The positions of the 221 GRBs in our sample are plotted in Fig.~\ref{fig:skycoverage1}, demonstrating fair sky coverage. Additionally, we plot the positions of low and high-$z$ GRBs separately in Fig.~\ref{fig:skycoverage2}. Both subgroups exhibit fair sky coverage, indicating no clustering of GRBs in any particular region of the sky.

\subsubsection{Model Selection using Akaike Information Criterion}
In this study, we employed two approaches to obtain the best-fit values of the Amati parameters: (1) a single linear regression applied over the entire redshift range, and (2) separate linear regressions for different redshift ranges. To compare the performance of these approaches, we utilized the Akaike Information Criterion (AIC) \citep{akaike1974new, akaike1987factor}, a widely used tool for model selection. For the single linear regression across the full redshift range, the AIC is defined as 

\begin{equation}
    AIC_{Full} = 2k - 2\log(L_{full}) \, ,
    \label{eq:aic}
\end{equation}
where $k$ represents the number of model parameters, and $L_{full}$ is the maximum likelihood of the linear model for the complete dataset. It is important to note that in the chi-squared statistic, $\chi^2$, introduced in Eq.~\ref{eq:chisq}, is additive. The likelihood function $L$ is proportional to $\exp{(-\chi^2 / 2)}$, and for independent observations, the total likelihood is obtained by multiplying the individual likelihoods. Accordingly, to compute the AIC for the subgroup analysis, we multiply the likelihoods for each redshift group, assuming the independence of observations for individual GRBs. The AIC for the grouped analysis is given by

\begin{equation}
    AIC_{Groups} = 2k - 2\log(L_{low-z} \times L_{high-z}) \, ,
\end{equation}
where $L_{low-z}$ and $L_{high-z}$ are the likelihoods of the linear models for the low-redshift and high-redshift subgroups, respectively. Lower AIC values indicate a better-fitting model by penalizing models with excessive complexity, thereby reducing the risk of overfitting. To quantify the difference between the models, we define $\Delta AIC$ as

\begin{equation}
    \Delta AIC = AIC_{Full} - AIC_{Groups} \, .
    \label{eq:deltaaic}
\end{equation}
The $\Delta AIC$ value helps determine whether the compared models differ significantly in terms of their fit to the data. If $\Delta AIC \geq 2$, the difference is considered significant, with the model having the lower AIC value being preferred. 
\section{Results and Discussion}
\label{sec:resndisc}
In this section, we present our main results and highlight the possible instrumental biases that could be responsible for any differences between the low and high-$z$ GRB properties. Toward the end of the present section, we also discuss the main differences in the properties of host galaxies of low and high-$z$ long GRBs.
\begin{table}
\begin{center}
\caption{Best-fit values of Amati parameters ($a$, $b$ and $\sigma_{int}$) along with their $1\sigma$ errors for data sample G ($221$ GRBs) and its subgroups $\rm{G_1}$ and $\rm{G_2}$. The cut-off has been taken at $z=1.5$ to divide the data into subgroups. The Bayesian marginalization has been employed to determine the best-fit value of each parameter.}
\label{tbl:bestfit-zeq1.5} \bigskip 
\begin{tabular}{cccc}
\hline
 Data Set & $a$ & $b$ & $\sigma_{int}$\\
\hline
G & $-6.30\pm 0.40$ & {$1.46\pm 0.07$} & $0.93\pm 0.05$ \\
$\rm{G_1(100)}$ & $-7.05\pm 0.55$ & $1.56\pm 0.09$ & $0.95\pm 0.08$ \\
$\rm{G_2(121)}$ & $-4.20\pm 0.70$ & $1.14\pm 0.12$ & $0.88\pm 0.07$ \\

\hline\hline
\end{tabular}
\end{center}
\end{table}
\begin{figure*}
  \begin{minipage}[t]{.33\linewidth}
\includegraphics[width=\linewidth]{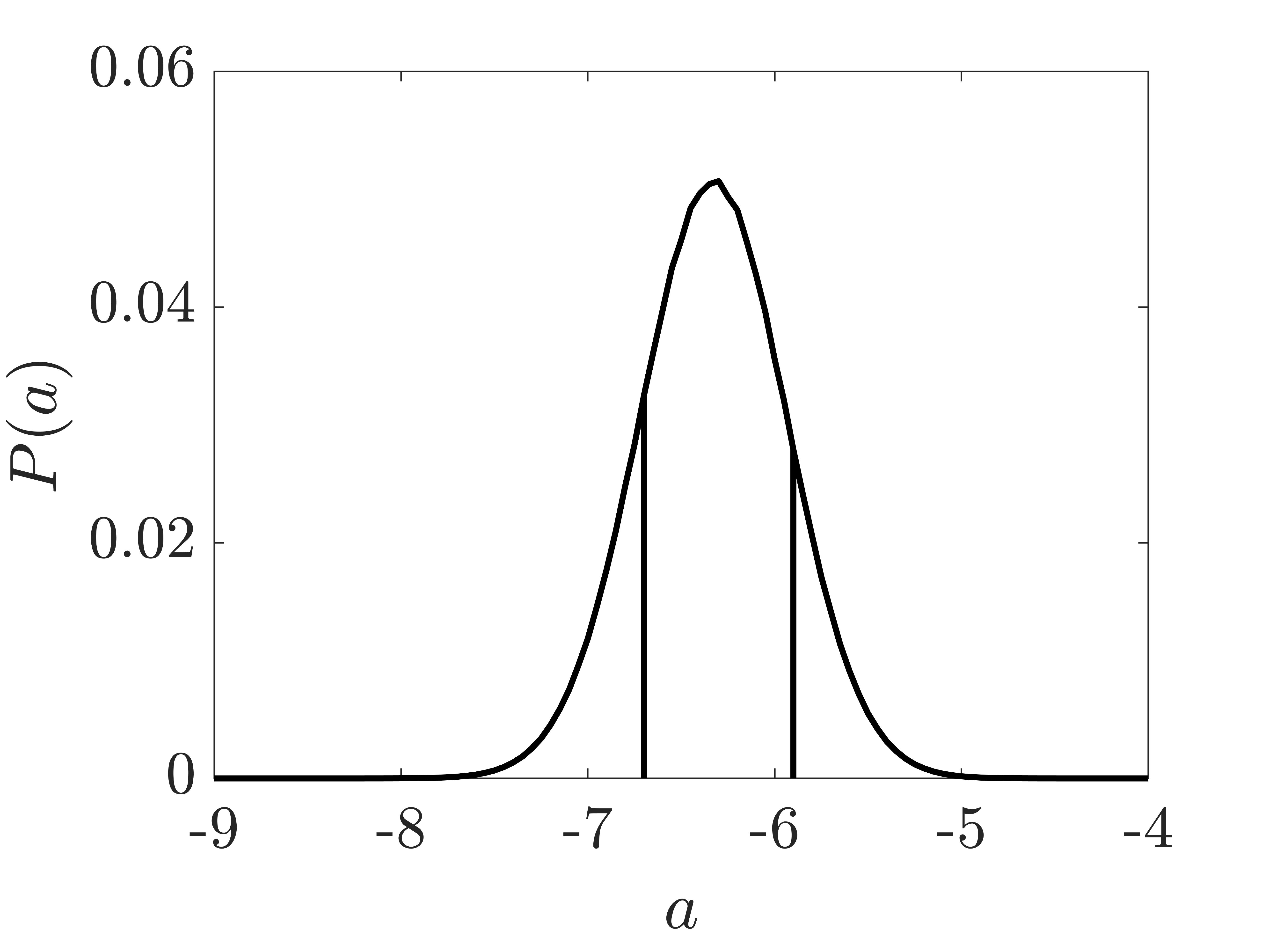}%
    \subcaption{}%
  \end{minipage}\hfil
  \begin{minipage}[t]{.33\linewidth} \includegraphics[width=\linewidth]{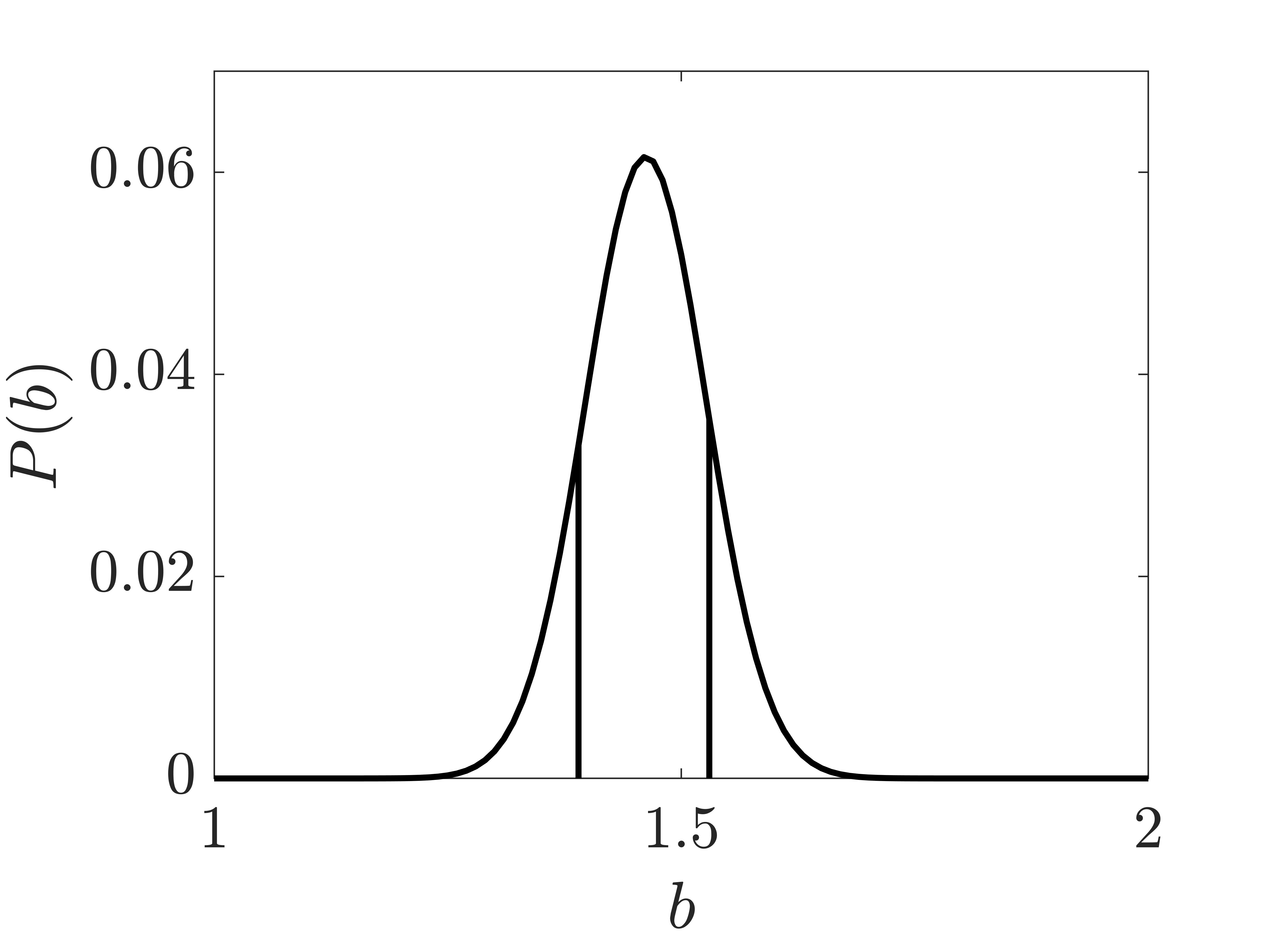}%
    \subcaption{}
  \end{minipage}\hfil
  \begin{minipage}[t]{.33\linewidth}
\includegraphics[width=\linewidth]{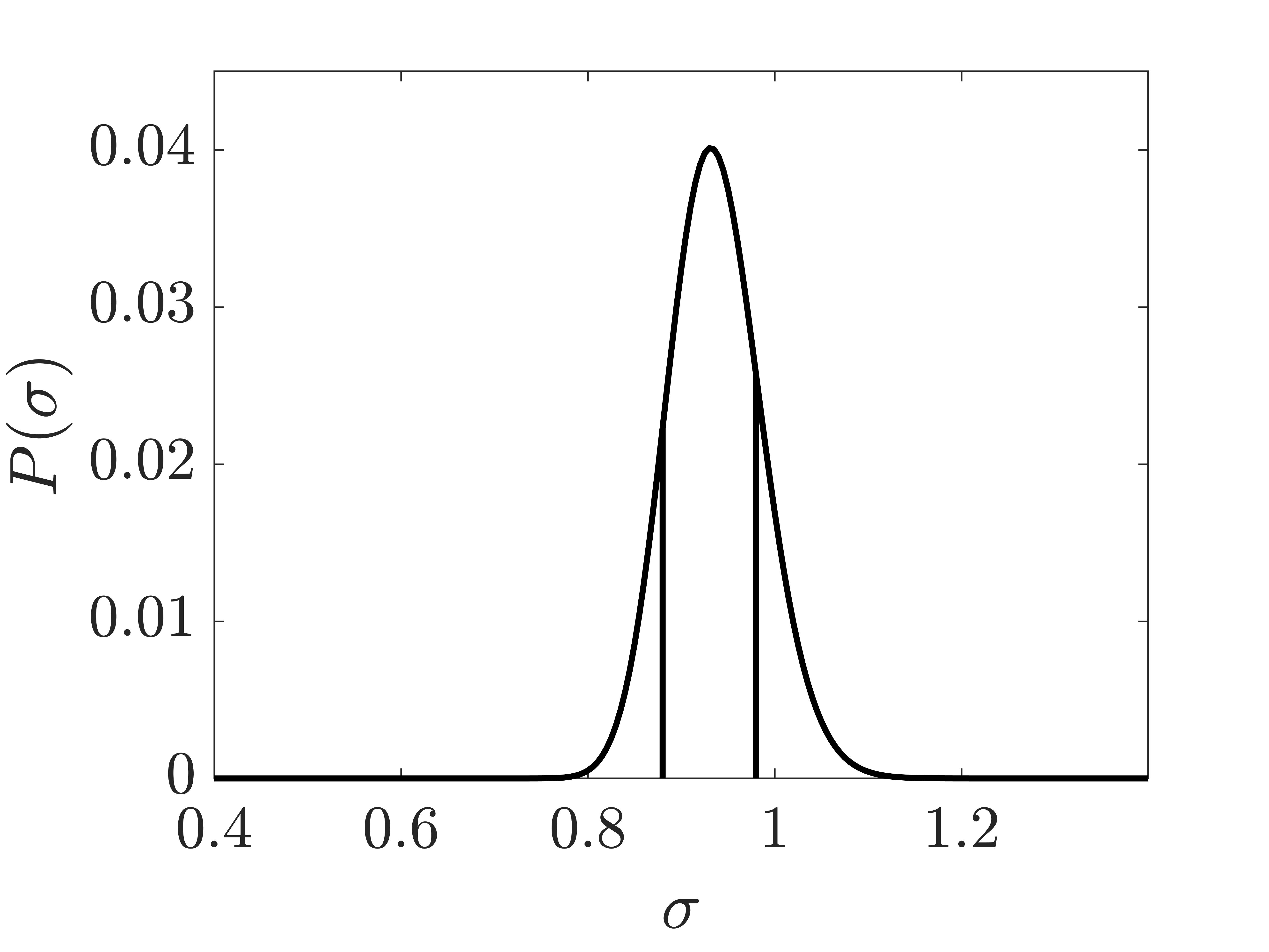}%
    \subcaption{}
  \end{minipage}%
  \caption{Distribution of posterior probability for Amati parameters $a$, $b$ and $\sigma_{int}$ for data sample G (221GRBs).}
  \label{fig:bf-221}
\end{figure*}

\begin{figure*}
  \begin{minipage}[t]{.33\linewidth}
    \includegraphics[width=\linewidth]{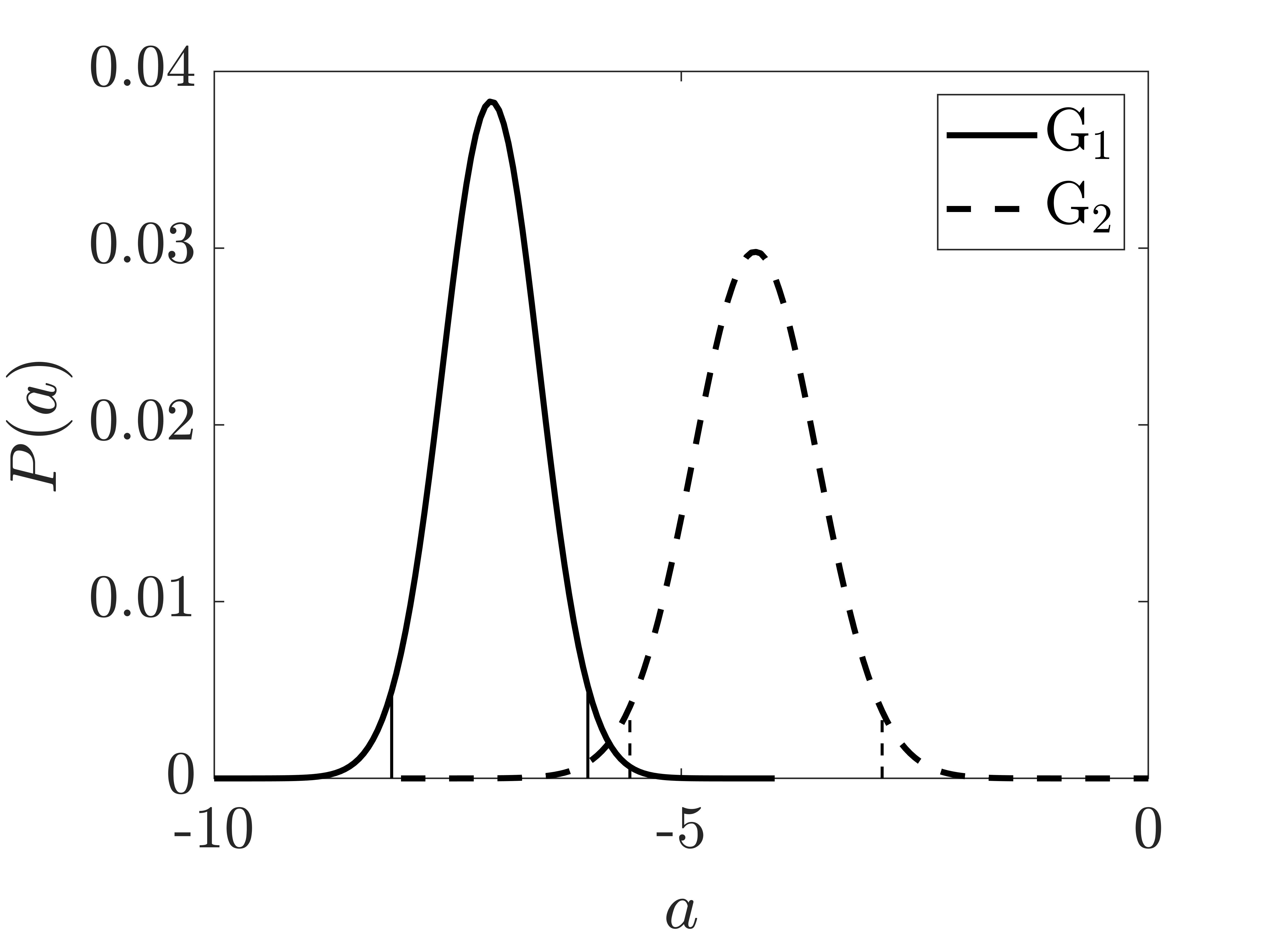}%
    \subcaption{}%
  \end{minipage}\hfil
  \begin{minipage}[t]{.33\linewidth}
    \includegraphics[width=\linewidth]{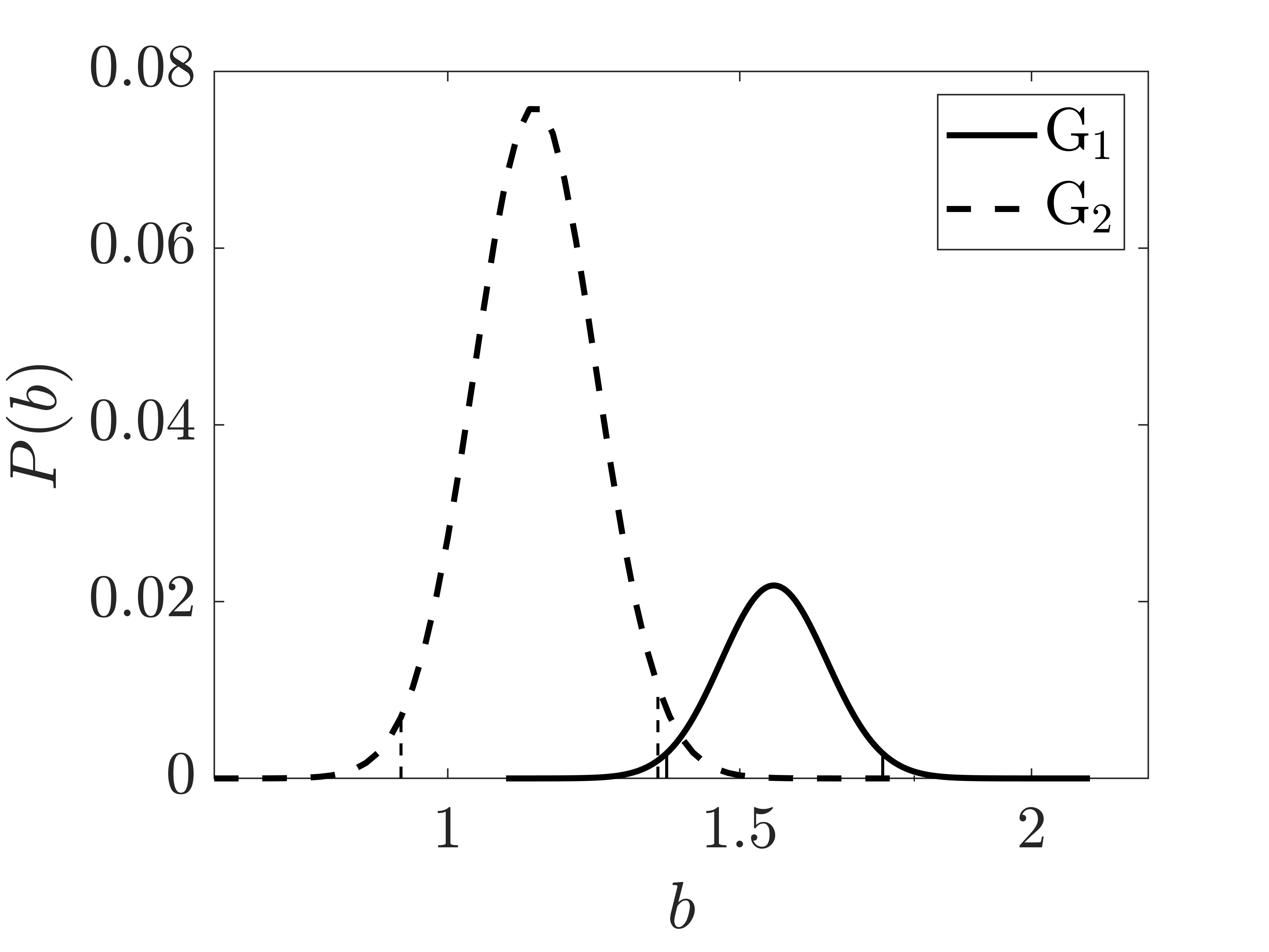}%
    \subcaption{}%
  \end{minipage}\hfil
  \begin{minipage}[t]{.33\linewidth}
    \includegraphics[width=\linewidth]{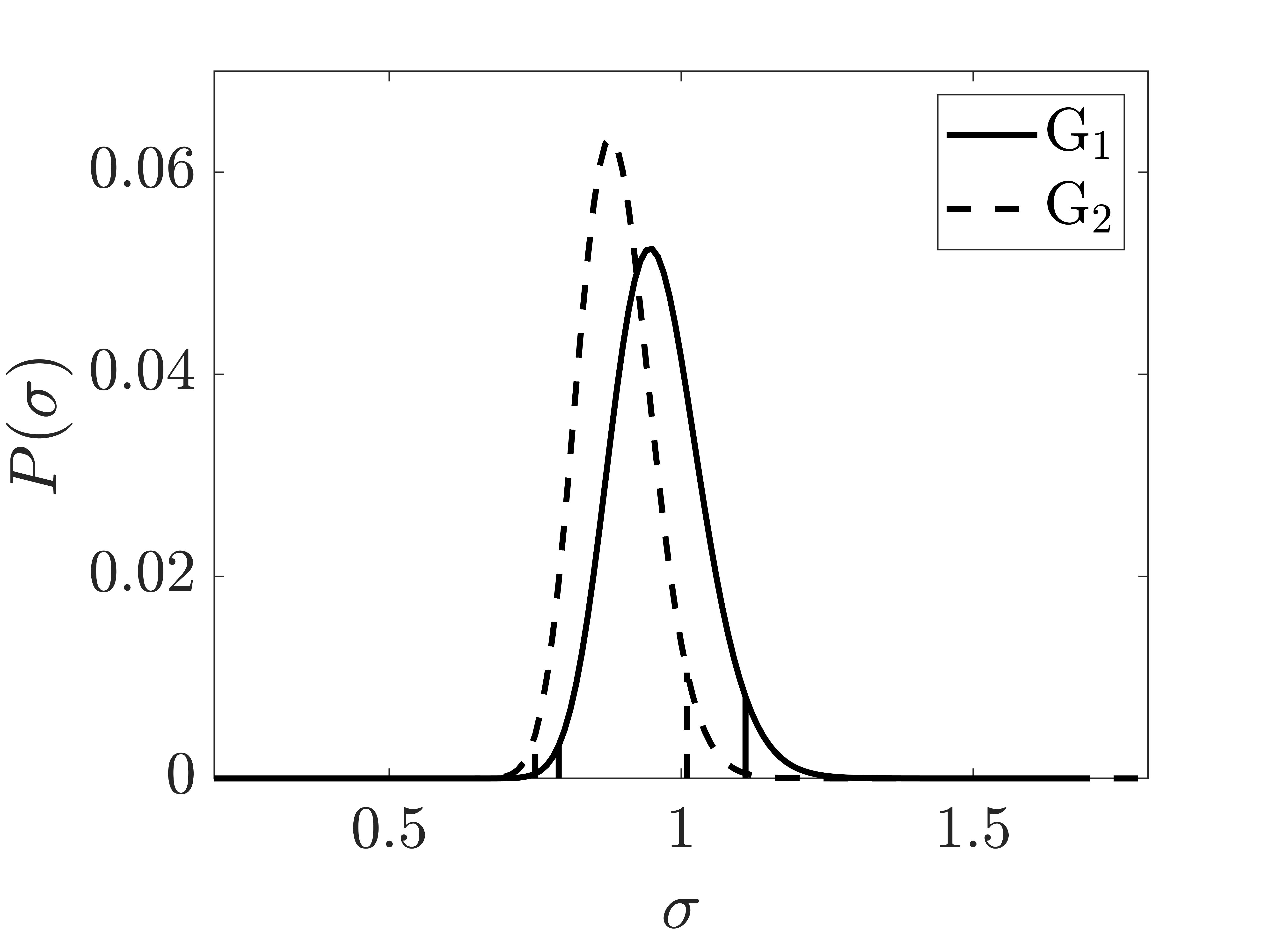}%
    \subcaption{}%
  \end{minipage}%
  \caption{Posterior probability distributions of Amati parameters $a$, $b$ and $\sigma_{int}$ for low-$z$ GRBs (group $\rm{G_1}$) and high-$z$ GRBs (group $\rm{G_2}$) with the threshold at $z=1.5$. The low-$z$ and high-$z$ distributions are not in agreement and their difference is statistically significant at more than $2 \sigma$ confidence level for parameters $a$ and $b$.
}
  \label{fig:G1-G2}
\end{figure*}
\begin{table}
\begin{center}
\caption{Best-fit values of $a$, $b$ and $\sigma_{int}$ along with their $1\sigma$ errors for the subgroups $\rm{G'_1}$ and $\rm{G'_2}$. The cut-off has been taken at $z=2$ to divide the data into subgroups. The Bayesian marginalization has been employed to determine the best-fit value of each parameter.}
\label{tbl:bestfit-zeq2} \bigskip 
\begin{tabular}{cccc}
\hline
 Data Set & $a$ & $b$ & $\sigma_{int}$\\
\hline
$\rm{G'_1}$ & $-6.75\pm 0.50$ & $1.52\pm 0.09$ & $0.99\pm 0.07$ \\
$\rm{G'_2}$ & $-4.60\pm 0.75$ & $1.21\pm 0.12$ & $0.80\pm 0.07$ \\
\hline\hline
\end{tabular}
\end{center}
\end{table}
\begin{figure*}
  \begin{minipage}[t]{.33\linewidth}
    \includegraphics[width=\linewidth]{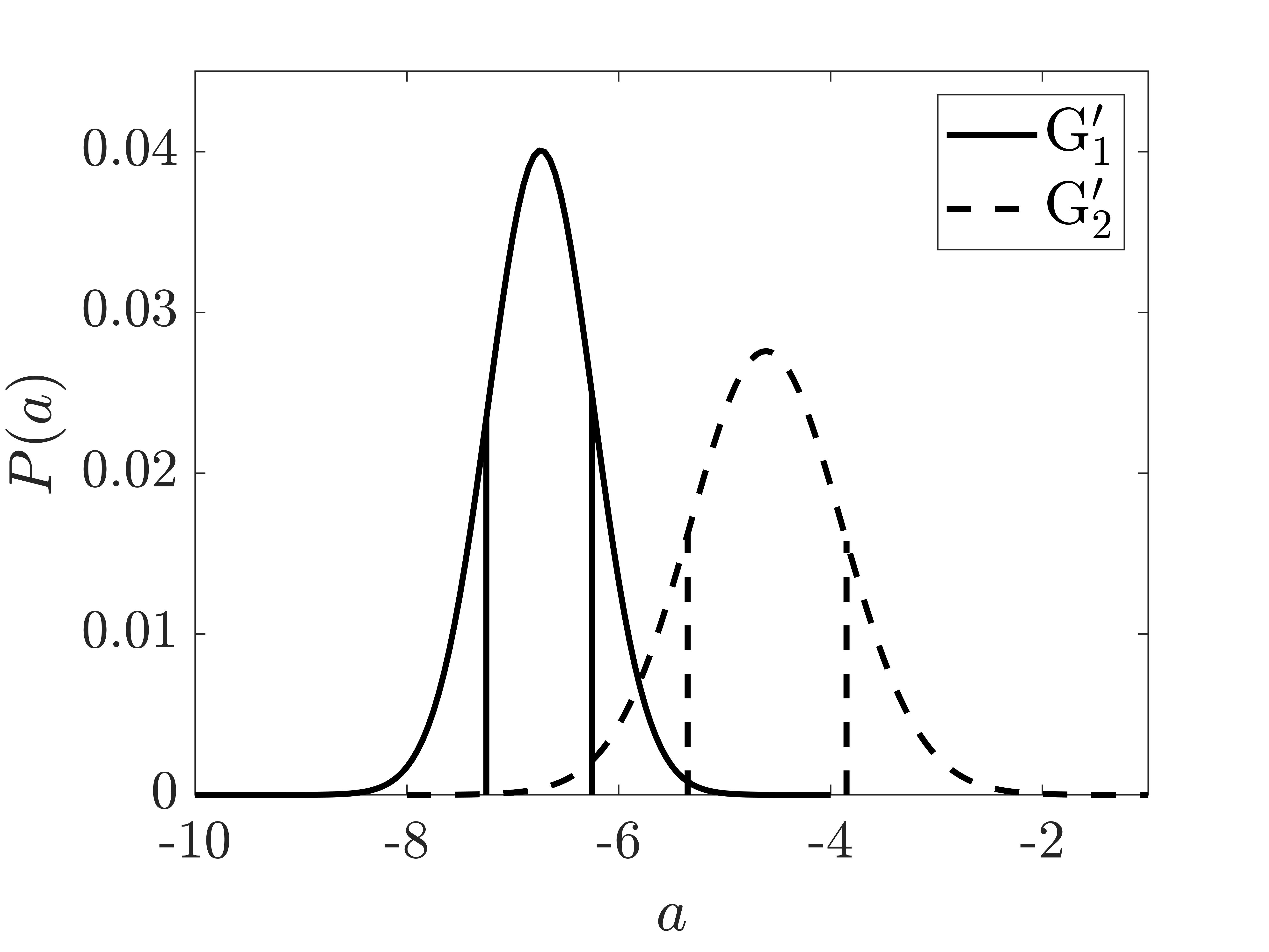}%
    \subcaption{}%
  \end{minipage}\hfil
  \begin{minipage}[t]{.33\linewidth}
    \includegraphics[width=\linewidth]{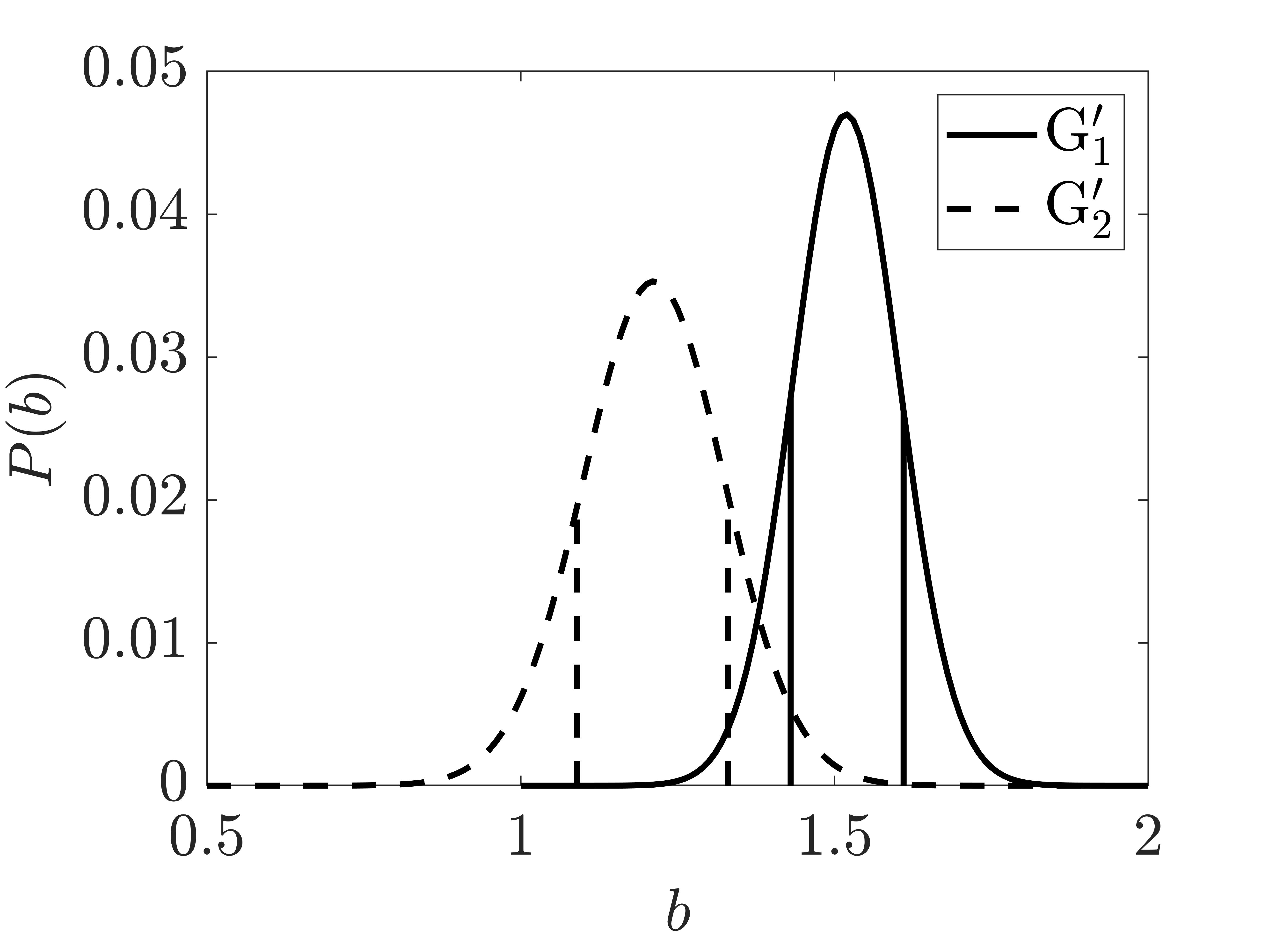}
    \subcaption{}%
  \end{minipage}\hfil
  \begin{minipage}[t]{.33\linewidth}
    \includegraphics[width=\linewidth]{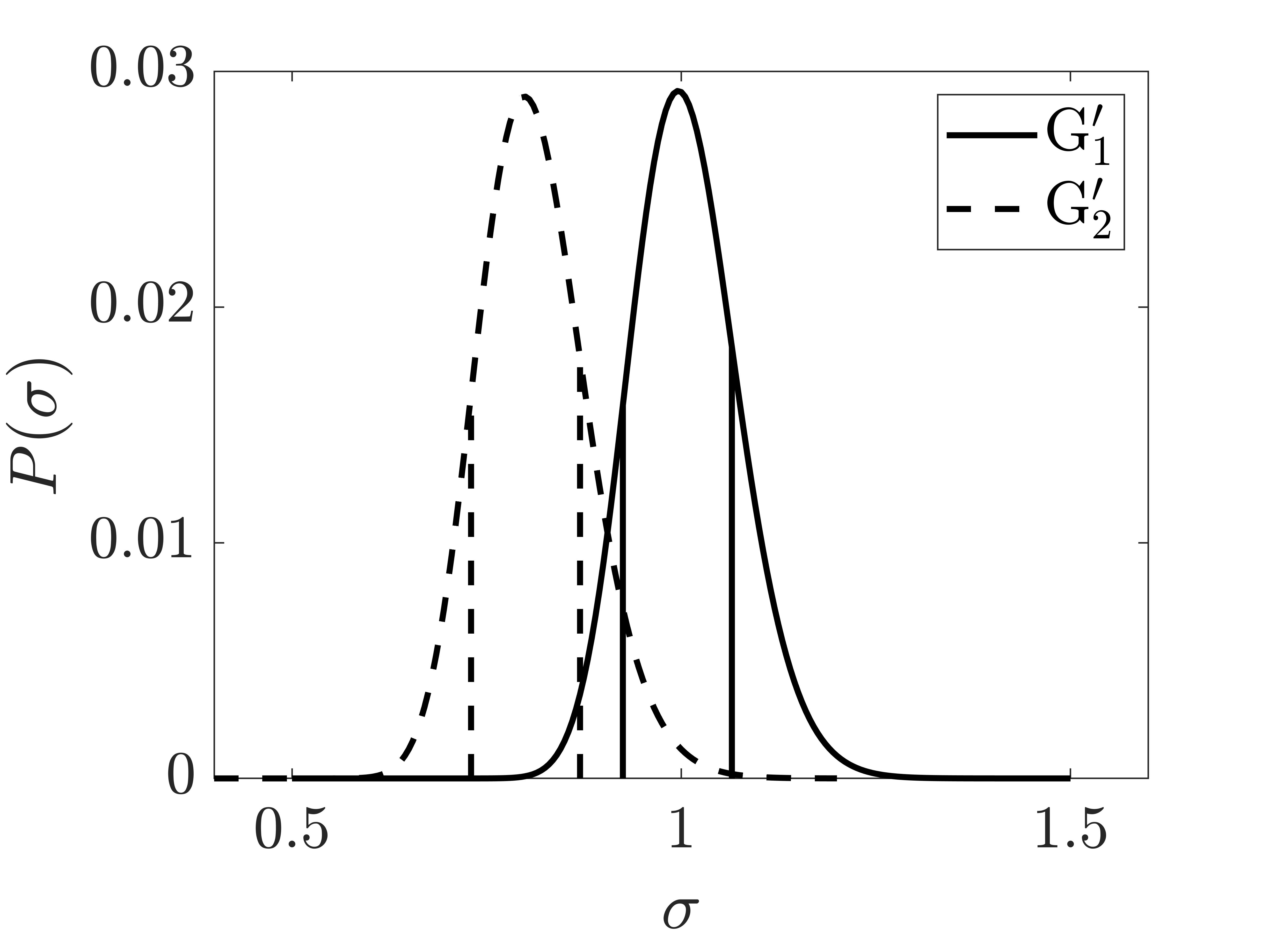}%
    \subcaption{}%
  \end{minipage}%
  \caption{Posterior probability distributions of Amati parameters $a$, $b$ and $\sigma_{int}$ for low-$z$ GRBs (group $\rm{G'_1}$) and high-$z$ GRBs (group $\rm{G'_2}$) with the threshold at $z=2$. The low-$z$ and high-$z$ distributions are not in agreement and their difference is statistically significant at more than $1 \sigma$ confidence level for each parameter.} 
  \label{fig:G'1-G'2}
\end{figure*}
\subsection{Main Results}
\label{sec:results}
The optimal parameters of the Amati relation are derived through a Bayesian approach, yielding the best-fit values and their associated $1\sigma$ uncertainties for the comprehensive dataset G, as summarized in Table~\ref{tbl:bestfit-zeq1.5}. Notably, the intercept ($a$) of the linear Amati relation exhibits a negative value, while its slope demonstrates a positive trend. The intrinsic scatter remains consistently below unity across all instances. Figure~\ref{fig:bf-221} visually represents the probability distribution of the three parameters, with vertical lines denoting the $1\sigma$ confidence level.

Table~\ref{tbl:bestfit-zeq1.5} also presents the optimal parameters, namely $a$ and $b$, for the distinct subgroups $\rm{G_1}$ and $\rm{G_2}$ (with the threshold at $z=1.5$). The intercept is still negative and the slope is positive in both subgroups. Figure~\ref{fig:G1-G2} illustrates the probability distributions associated with these subgroups, delineating the distinctive nature of $\rm{G_1}$ and $\rm{G_2}$. Vertical lines are employed to mark the $2\sigma$ confidence level. Notably, the values of $a$ and $b$ exhibit a discrepancy exceeding the $2\sigma$ threshold, indicating a discernible distinction in the GRB populations at low and high redshifts. Despite a smaller value observed at high redshifts, the disparity in the intrinsic scatter is not statistically significant.

\begin{figure*}
    \begin{minipage}[t]{.50\linewidth}
    \includegraphics[width=\linewidth]{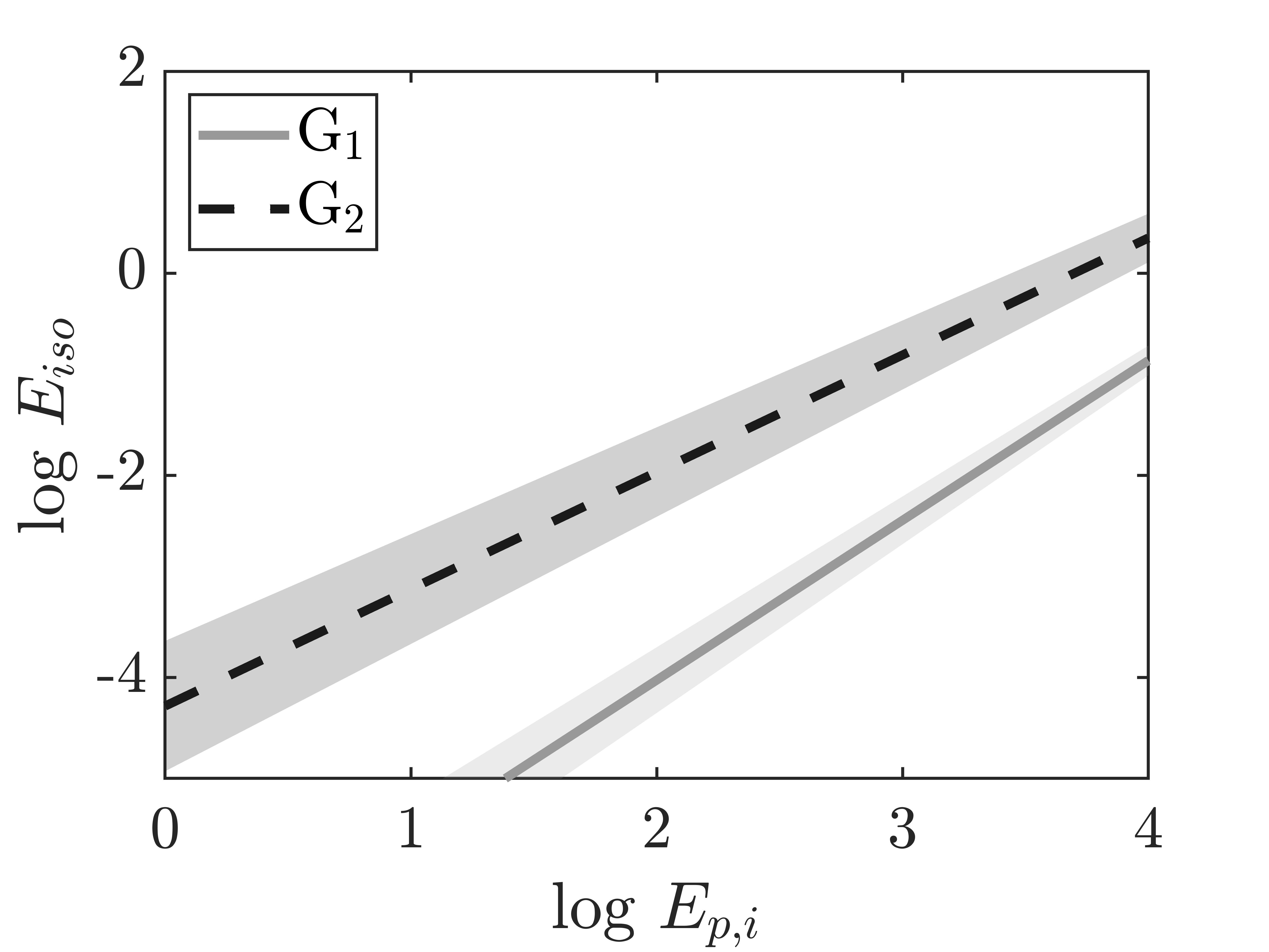}%
\subcaption{}
  \end{minipage}\hfil
  \begin{minipage}[t]{.50\linewidth}
 \includegraphics[width=\linewidth]{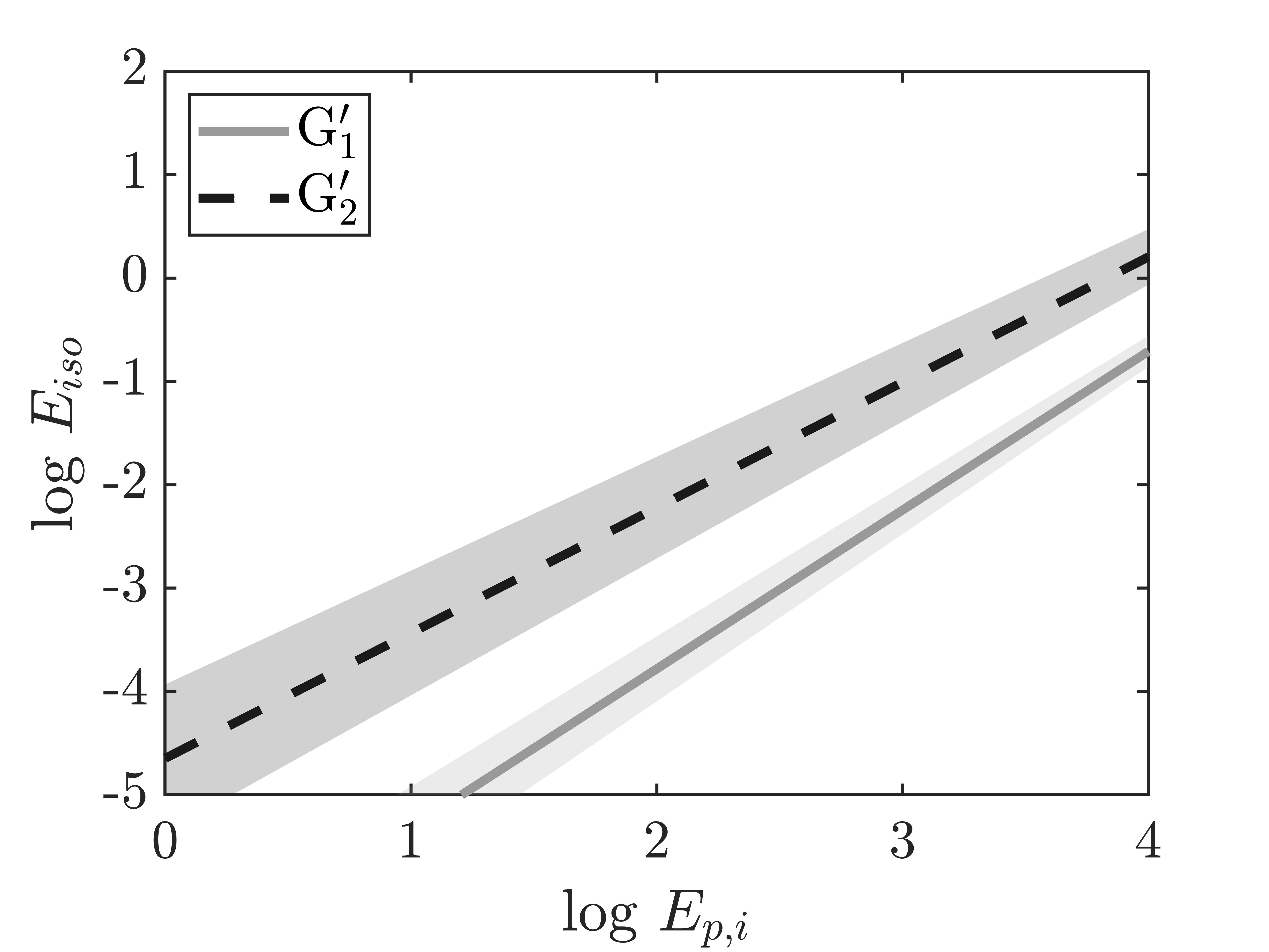}%
   \subcaption{}%
  \end{minipage}%
  \caption{Straight line fit $ y = a + bx $ for the subgroups $\rm{G_1}$, $\rm{G_2}$ in the left panel and for the subgroups $\rm{G'_1}$, $\rm{G'_2}$ in the right panel, where $a$ and $b$ are the best-fit values of Amati parameters.}
  \label{fig:linefit_1.5_2}
\end{figure*}

Table~\ref{tbl:bestfit-zeq2} presents the optimal values of $a$ and $b$, for the subgroups $\rm{G'_1}$ and $\rm{G'_2}$ (with the threshold at $z=2$). Their $1\sigma$ errors have also been shown in the respective columns. The continuously negative intercept and positive slope observed in both subgroups underscore a consistent trend. The corresponding posterior probability distributions for these parameters have been shown in Figure~\ref{fig:G'1-G'2}. The vertical lines have been drawn at $1\sigma$ level. The significance is comparatively smaller in this case however, it is still at more than $1\sigma$ for both $a$ and $b$. In this case, the intrinsic scatter is also significantly different for the two subgroups.
To illustrate the differences in the $E_{p,i} - E_{iso}$ relations between low-$z$ and high-$z$ GRBs, we present a comparative analysis in Figure~\ref{fig:linefit_1.5_2}. The left panel of Figure~\ref{fig:linefit_1.5_2} displays the straight line fits for subgroups $G_1$ and $G_2$, i.e., at the redshift threshold of $z = 1.5$. The solid line represents the fit for low-$z$ GRBs, while the dashed line corresponds to high-$z$ GRBs. The results indicate that the linear fits for low-$z$ and high-$z$ GRBs are distinct, with no overlap between the two groups. Similarly, the right panel presents the straight line fits for the subgroups $\rm{G'_1}$ and $\rm{G'_2}$. As in the previous case, both fits are distinct. 

To further validate our findings, we computed the AIC for the single linear fit applied across the entire redshift range and the fits for the individual subgroups. The AIC values are presented in Table~\ref{tbl:AIC}. The AIC for the single linear fit (full $z$ range) is the highest and lowest for subgroups $G_1$ and $G_2$. The double linear fit provides smaller AIC in both cases ($z_{crit} = 1.5$ and $2$) than the single linear fit for full data. The difference in AIC values, as defined in Eq.~\ref{eq:deltaaic}, exceeds the threshold for significance. Therefore, the subgroup-based fit at $z_{crit} =1.5$ is preferred, as it provides a better representation of the data than the single linear fit over the full redshift range.

\begin{table}[h!]
\centering
\caption{AIC comparison for single linear fit for full redshift range and double linear fits with breaks at $z=1.5$ and $z=2$.}
\label{tbl:AIC}
\begin{tabular}{lcc}
\hline
\text{Model} & \text{$z$ break} & \text{AIC} \\ 
\hline
Single Linear Fit (Full Redshift Range) & No & 220.98 \\ 
Double Linear Fit with break at $z = 1.5$ & Yes & 205.79 \\ 
Double Linear Fit with break at $z = 2.0$ & Yes & 212.73 \\ 
\hline
\end{tabular}
\end{table}

\subsection{Probable Instrumental Biases and Selection Effects}
Possible biases in each instrument could arise due to their design, calibration, and operational characteristics which may be partly responsible for the observed differences in GRB populations. BAT  may exhibit biases related to its coded aperture imaging technique, potentially leading to systematic errors in position determination \citep{Barthelmy_2005}. It can affect the identification of the host galaxy and hence the redshift measurement. Another issue with the BAT instrument is its capability to detect energies up to 150 keV only \citep{Gehrels2004}, which is less than the average peak energy of GRBs \citep{Kaneko_2006}. Thus, for several Swift observed GRBs, it is not possible to directly determine the fluence and $E_{peak}$. 
FGST's LAT may face biases due to background noise, energy calibration uncertainties, and instrument response variations across its wide energy range. GBM's biases could include sensitivity variations across its energy range and potential systematic errors in event classification. While these instruments provide invaluable data, understanding and mitigating biases are crucial for accurate gamma-ray observations.

\subsection{Low-$z$ vs High-$z$ GRB Hosts}
As mentioned in \ref{sec:results}, we observe significant differences in best-fit values of Amati parameters for low and high-$z$ GRB subgroups. Below we analyse our results in the context of differences in the low-$z$ and high-$z$ GRB host galaxies. 

In a detailed study \citep{Palla2020} and \citep{Grieco_2014} claim that at high $z$, long GRBs host galaxies include all morphological types including early-type galaxies. This could be because at high $z$ the early-type galaxies can be observed during their active star formation period. In contrast, star formation in the early-type galaxies at low $z$ is often quenched.

High-$z$ GRB hosts are younger and thus have low metallicities. On the other hand, low-$z$ hosts, especially the early-type ones, are older and have higher metallicities \citep{Palla2020}. These differences underscore significant evolutionary changes in galaxy properties and environments from high to low redshifts, affecting the nature and identification of long GRB host galaxies. At a given redshift, dust-obscured GRB hosts are more massive than optically bright galaxies, with massive GRB hosts becoming fairly common at $z>1.5$ \citep{Perley_2013}. The authors also link the smaller GRB production at low redshift to metallicity.
Using the optically unbiassed GRB host (TOUGH) Survey \citep{Schulze_2015} claims that GRB hosts at high $z$ are faint and low in metallicity \citep{2015ApJ...804...51C}. 
The studies based on Damped Lyman $\alpha$ systems at high $z$ support the existence of two GRB progenitor channels, with one being mildly influenced by metallicity \citep{2015ApJ...804...51C}.

\section{Conclusions}
\label{sec:conclusion}

This study explored the potential heterogeneity in the Amati correlation among long gamma-ray bursts (GRBs) across varying redshift ranges. Building on earlier investigations by \citep{ghirlanda2010, Li2007, Singh2024, Wang_2017}, we addressed concerns regarding sample limitations and inconsistencies highlighted by \citep{Khadka2021} and \citep{Dirirsa2019}. 
By leveraging an expanded dataset of 221 long GRBs, spanning redshifts from 0.034 to 8.2, we conducted a comprehensive analysis to better understand the redshift dependence of the Amati relation. Our findings reveal statistically significant variations in the Amati relation parameters across different redshift subgroups. When dividing the dataset at critical redshifts, \(z_{\text{crit}} = 1.5\) and \(z_{\text{crit}} = 2.0\), the fitting parameters \(a\) and \(b\) showed noticeable differences. 
At \(z_{\text{crit}} = 2\), the low- and high-redshift subgroups exhibited a difference of more than \(1\sigma\). At \(z_{\text{crit}} = 1.5\), the differences exceeded \(2\sigma\) for $a$ and $b$ parameters, indicating a more pronounced distinction between low- and high-redshift GRB populations. 
These variations may be attributed to intrinsic factors such as host galaxy metallicity, although selection effects or instrumental biases could also play a role. Our findings align with those of \citep{Singh2024}, who reported similar discrepancies at \(z_{\text{crit}} = 1.5\) using a different dataset.
Furthermore, the Akaike Information Criterion (AIC) analysis favored a double-line fit with a break at \(z = 1.5\) over a single-line fit, reinforcing the idea that GRBs may exhibit distinct properties across redshift ranges. Looking ahead, upcoming missions such as the \textit{Transient High-Energy Sky and Early Universe Surveyor} (THESEUS), slated for launch in 2032 \citep{2021ExA....52..277G}, and the \textit{enhanced X-ray Timing and Polarimetry} (eXTP) mission \citep{Zhang_2019}, promise to advance our understanding of high-redshift GRBs. THESEUS aims to detect GRBs up to \(z = 12\), while eXTP is expected to observe approximately 100 GRBs annually, offering enhanced sensitivity and broader observational coverage. These missions will provide the precision and data volume necessary to validate our results and further clarify the differences between low- and high-redshift GRB populations.

\section*{Acknowledgments}
Meghendra Singh thanks DMRC for support. Darshan Singh thanks the compeers of GD Goenka University for eternal assistance. 

 \bibliographystyle{elsarticle-num} 
 \bibliography{main}



\end{document}